\documentclass[%
 reprint,
 amsmath,amssymb,
 aip,
 apl,
]{revtex4-2}

\usepackage{graphicx}% Include figure files
\usepackage{dcolumn}% Align table columns on decimal point
\usepackage{bm}% bold math
\usepackage{setspace}
\usepackage{hyperref}
\begin{document}

\preprint{APS/123-QED}

\title{Electronic and optical properties of 4H Si from first principles}% Force line breaks with \\

\author{Xiao Zhang}
\author{Emmanouil Kioupakis}
\affiliation{Department of Materials Science and Engineering, University of Michigan, Ann Arbor, Michigan 48109, USA}
\date{\today}

\begin{abstract}
 The cubic polytype of silicon (Si) is the most commercialized semiconductor material and finds applications in numerous electronic and optoelectronic devices such as solar cells. However, recent reports on the synthesis of the hexagonal 4H Si polytype have attracted the attention of the scientific community to understand its functional properties. Here we report the electronic, vibrational, and optical properties of the 4H Si polytype obtained with predictive first-principles calculations. We find that, compared to the cubic polytype, 4H Si shows a slightly narrower indirect gap by $\sim$ 0.05 eV. By calculating its direct and phonon-assisted optical spectra, we show that 4H Si exhibits a stronger absorption coefficient than cubic Si across the visible and IR spectral regions. We further evaluate the short-circuit current density of textured thin-films, and we demonstrate that 4H Si can achieve the same short-circuit current density for a five times thinner film compared to the cubic polytype. Our work demonstrates the advantages of 4H Si for thin-film silicon-based solar-cell applications. 
 \end{abstract}

\maketitle

%\tableofcontents

Silicon (Si) is a semiconducting material widely used in optoelectronic devices and is by far the most widely deployed solar-cell technology. The properties of the conventional cubic Si have been thoroughly studied both from an experimental and from a theoretical point of view. While there are many aspects that drove Si to be the dominant semiconductor for solar cells, the indirect nature of its band gap (with a magnitude of 1.1 eV at room-temperature) limits the capability of Si for thin-film solar applications. Simulations have shown that the optimum thickness of Si solar cell is on the order of 100 µm\cite{tiedje1984}, in contrast to many thin-film solar cells whose thickness can be on the order of a few microns or even in the nano-region\cite{lee2017review,powalla2018thin,nakamura2019cd}. Such a limitation for cubic Si is fundamental, and in order to overcome it in silicon-based solar cells, it is necessary to explore other possible routes, such as amorphous Si\cite{ru202025,qarony2017efficient} or different Si polytypes.

Recently, the hexagonal structure of Si (4H) has been successfully synthesized\cite{PhysRevLett.126.215701}, which provides a promising new route to improve the efficiency of silicon-based solar cells. Experimentally, it has been shown that bulk 4H Si can be synthesized through metastable phase transformation upon heating the single-crystalline Si$_{24}$ allotrope\cite{PhysRevLett.126.215701}. In addition, experimental studies also find that 4H Si exhibits similar mechanical properties compared to the cubic polytype, and can be potentially integrated in photovoltaic applications\cite{liang2022mechanical}. Previous theoretical calculations on found that the fundamental indirect band gap of 4H Si can be expected to be narrower than its cubic polytype by 0.05-0.1 eV\cite{persson1998electronic,raffy2002properties}. Moreover, due to symmetry breaking in the hexagonal structure compared to the cubic structure, stronger electron-phonon interaction are expected in the hexagonal form, thus enhancement of the phonon-assisted absorption in the indirect regime can potentially lead to enhancement in the performance of optoelectronic devices such as solar cells. To understand the potential of the 4H Si polytype in optoelectronic applications such as solar cells, it is essential to understand its optical absorption spectrum especially in the indirect regime. 

In this work, we present a first-principles evaluation of the structural, electronic, and optical properties of 4H Si, including its indirect optical properties assisted by electron-phonon interactions. We show that our theoretically calculated structural and electronic properties agree well with previous studies. Using the $GW$ approximation, we calculate the quasiparticle band structure of both hexagonal and cubic polytypes, and find that the 4H polytype exhibits both a narrower indirect band gap by 0.05 eV as well as a much narrower direct band gap by 1.46 eV. The direct absorption spectra of 4H Si show that it exhibits a much lower absorption onset compared to the cubic polytype. We then calculate the absorption coefficient of 4H Si and demonstrate that the 4H Si is a better absorber compared to cubic Si in the entire spectral range of IR and visible. Further analysis on solar-cell performance shows that the 4H polytype achieves the same short-circuit current density as the cubic polytype for a five times lower thickness. Our computational study demonstrates that 4H Si is a promising candidate for increasing the efficiency of silicon-based solar cells.

Our first-principles methods for calculating the structural, electronic, and optical properties of 4H Si are based on density function theory\cite{PhysRev.140.A1133, PhysRev.136.B864} and related techniques. Ground-state calculations and relaxations are performed using the Quantum Espresso\cite{giannozzi2009quantum,giannozzi2017advanced} package using the SG15 Optimized Norm-Conserving Vanderbilt (ONCV) pseudopotentials\cite{schlipf2015optimization,PhysRevB.88.085117} with the Perdew-Burke-Ernzerhof (PBE) exchange-correlation funtional\cite{PhysRevLett.77.3865}
The relaxed lattice constant is obtained via fitting to the energy-volume curve to the Murnaghan equation of state\cite{EoS}. 
To obtain quasiparticle energies, we adopted the single-shot GW approximation as implemented in the BerkeleyGW code\cite{DESLIPPE20121269,hybertsen1986}.
To calculate direct optical properties, we solve the Bethe-Salpeter equation for optical polarization functions to take electron-hole Coulomb interactions into account\cite{rohlfing2000o}. 
The quasiparticle energies are interpolated using the Maximally Localized Wannier Function (MLWF) approach\cite{RevModPhys.84.1419} to obtain the electronic energies on a finer grid for band structures and phonon-assisted optical property calculations. 
Indirect optical absorption is characterized by second-order perturbation theory\cite{Noffsinger2012,zhang2022} and implemented in the EPW code\cite{Ponce2016, Giustino2007,lee_electronphonon_2023}.

\begin{figure}
    \centering
    \includegraphics[width=0.4\textwidth]{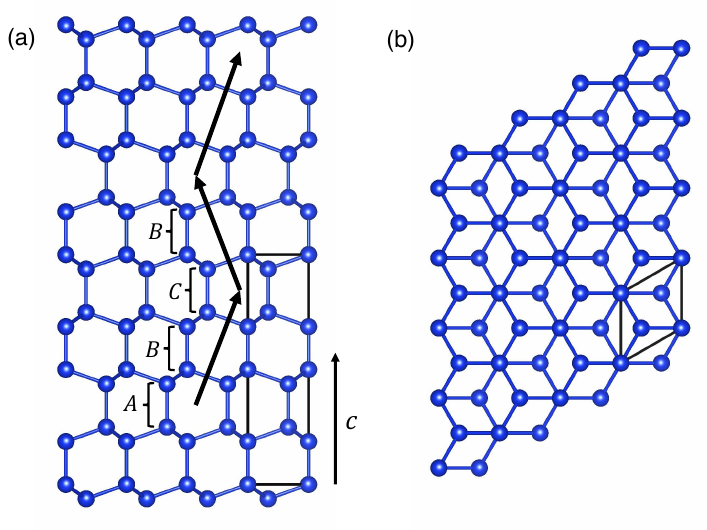}
    \caption{The structure of 4H Si viewed from (a) the $[11\bar{2}0]$ direction and (b) the [0001] (crystalline $c$) direction. The black box marks the simulation unit cell. The $ABCB$ stacking sequence along the crystalline $c$ direction is clearly seen in (a).}
    \label{fig:structure}
    \vspace{-0.5cm}
\end{figure}

We first show that our calculated lattice parameters of the 4H polytype agree well with both experimental and theoretical studies. Figure~\ref{fig:structure} demonstrates the crystal structure of the 4H Si polytype. The structure is similar to the 4H SiC polytype with an $ABCB$-type stacking pattern\cite{morkoc1994} along the [0001] direction. The fully relaxed lattice parameters of 4H Si from our simulations are $a$ = 3.86 \r{A}, $c$ = 12.69 \r{A}. Our results overestimate experimental measurements by 0.5\% for the $a$ and 0.8\% for the $c$ lattice constant. This is expected due to the underbinding nature of the PBE exchange-correlation functional. The difference of the lattice parameters compared to the experimental measurements is also found to be smaller than previous studies that used the LDA exchange-correlation functional\cite{persson1998electronic,raffy2002properties}. Our fully relaxed lattice constant for cubic Si is $a$ = 5.478 Å, which is in good agreement with previous theoretical studies\cite{mo2018accurate} and shows the same trend of overestimating the experimentally measured\cite{omara2007o} value of 5.431 \r{A} by 0.8\%.

\begin{figure}
    \includegraphics[width=0.4\textwidth]{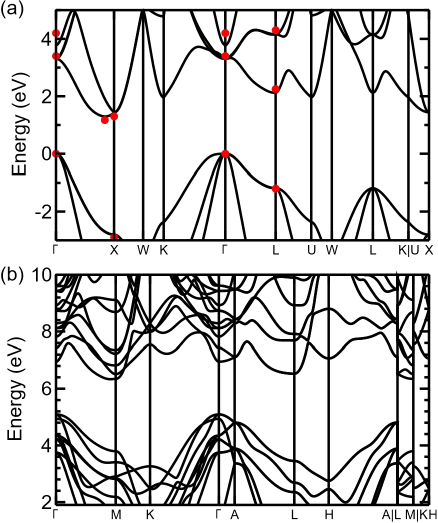}
    \caption{Electronic band structure of (a) cubic Si and (b) 4H Si. The calculated band gaps of the two polytypes are both indirect: E$_g$ = 1.29 eV for cubic and E$_g$ = 1.24 eV for 4H. Our calculated band structure of cubic Si is in good agreement with experimental measurements (Ref.\onlinecite{hybertsen1986} and references therein). The minimum direct gap of 4H Si (1.85 eV at $\Gamma$) is much smaller compared to the cubic polytype (3.31 eV at $\Gamma$).}
    \label{fig:bands}
    \vspace{-0.5cm}
\end{figure}

We next calculate the band structures for both 4H and cubic Si (Fig.~\ref{fig:bands}). The calculated quasiparticle energies for the cubic polytype agree well with experimental measurements\cite{hybertsen1986}.
Our calculated band gap of 4H Si (E$_g$=1.24 eV) agrees well with the original report of the bulk 4H Si structure (E$_g$=1.22 eV, extracted from the calculated band structure obtained with the HSE06 exchange-correlation functional in Ref. \onlinecite{PhysRevLett.126.215701}). Comparing the two structures, we find that the band gap of 4H Si is 0.05 eV narrower than that of the cubic structure (1.29 eV). 
The difference of the band gap between the two polytypes is also in good agreement with previous predictions\cite{persson1998electronic,raffy2002properties}. 
For cubic Si, the global minimum of the conduction bands is located at a point at approximately 85\% of the $\Gamma-X$ direction, while the global minimum of the conduction bands for 4H Si is located close to the $M$ point. 
Notably, the minimum direct gap of 4H Si (1.85 eV) is also significantly smaller than that of the cubic Si (3.31 eV). 
The lower onset of direct absorption of 4H Si makes it a more promising option for thin-film solar cells compared to cubic Si.

\begin{figure}
    \centering
    \includegraphics[width=0.4\textwidth]{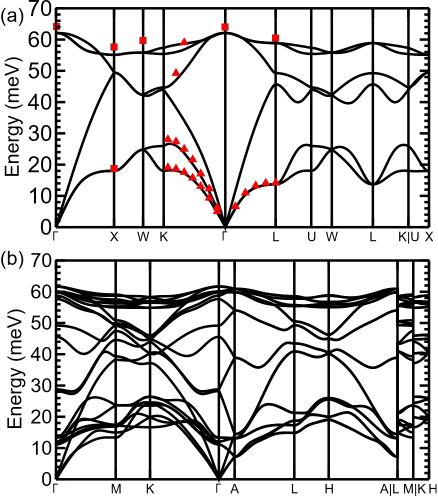}
    \caption{Calculated phonon dispersion relation of (a) cubic Si and (b) 4H Si. The calculated maximum phonon frequency of the two polytypes are very close: 62.1 meV (cubic) versus 61.7 meV (4H). Our calculated phonon dispersion for cubic Si agrees within 1\% compared to Ref.\onlinecite{petretto2018high} and is in good agreement with experimental measurements at high-symmetry points from Ref.\onlinecite{PhysRevB.6.3777} (red rectangles) and Ref.\onlinecite{dolling1963} (red triangles).}
    \label{fig:phbands}
    \vspace{-0.5 cm}
\end{figure}

Since 4H Si is an indirect-gap material, like cubic Si, it is necessary to consider electron-phonon interactions in order to determine its indirect optical absorption spectrum. We plot the calculated phonon dispersion relations of both the cubic and the 4H polytype in Fig.~\ref{fig:phbands}. We find very good agreement with phonon dispersion for cubic Si calculated using ABINIT\cite{petretto2018high}, with a difference of less than 1\% for the maximum phonon frequency at $\Gamma$. Our calculated phonon frequencies for cubic Si at high-symmetry points of the BZ are also in good agreement with experimental measurements\cite{dolling1963,PhysRevB.6.3777} for the acoustic modes and show an underestimation of less than 3\% for the optical modes. Overall, we see that the maximum phonon frequencies of the two polytypes are comparable (62.1 meV for cubic Si versus 61.7 meV for 4H Si). We further evaluate the sound velocity along the (100) direction. The calculated sound velocity for cubic Si is 5354 m/s and 8143 m/s for transverse and longitudinal modes, respectively, which agree well with the experimentally measured values of 5844 m/s and 8433 m/s\cite{mcskimin1964elastic}. For 4H Si, the sound velocity is 4960 m/s and 8607 m/s for transverse and longitudinal modes, respectively, i.e., $\sim$7\% lower for transverse mode and 3.5\% higher for the longitudinal mode.

\begin{figure}
    \centering
    \includegraphics[width=0.48\textwidth]{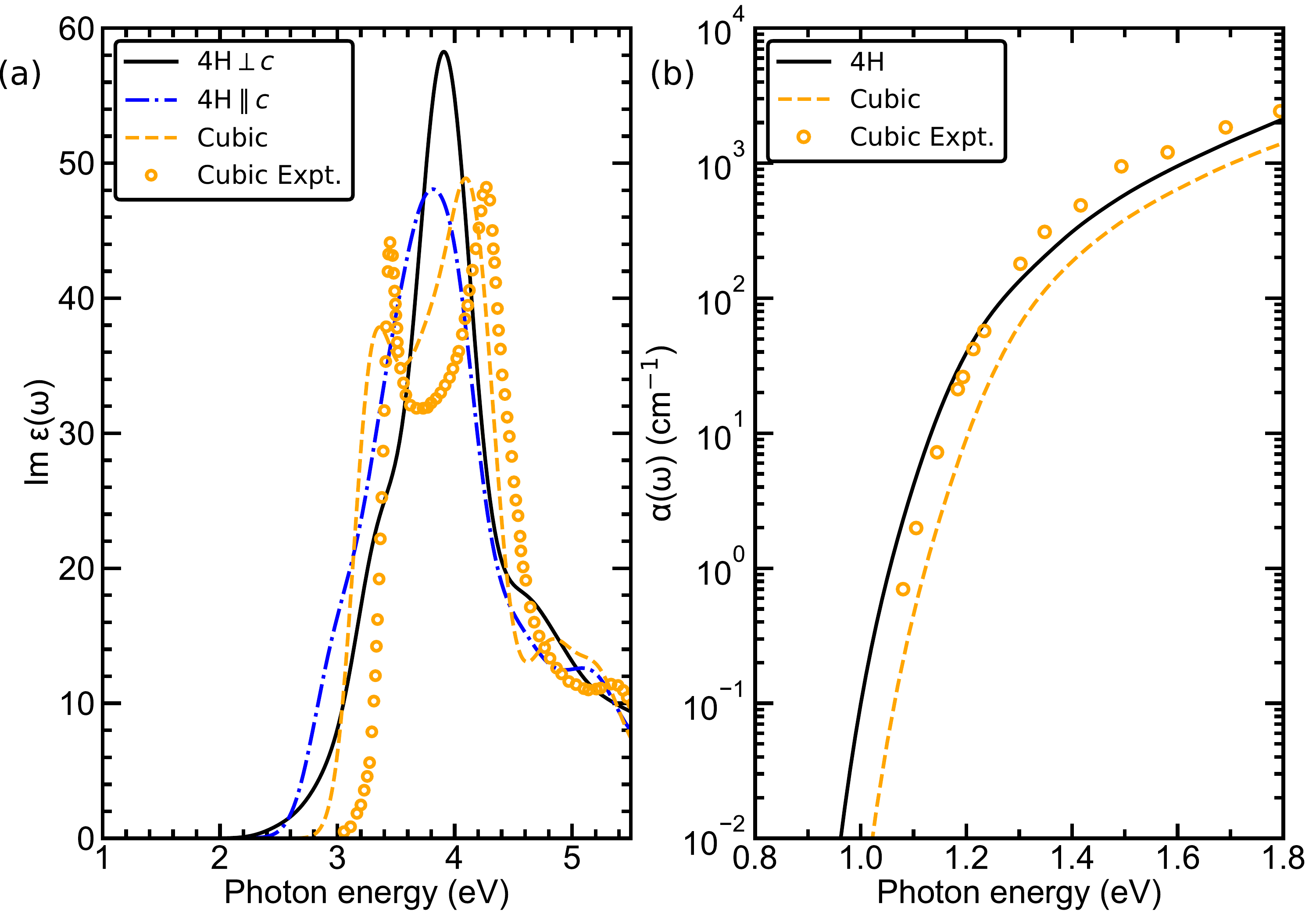}
    \caption{(a) Calculated direct optical absorption spectra (imaginary part of the dielectric function) of 4H Si for the ordinary ($E\perp c$, solid black) and the extraordinary ($E\parallel c$, dot-dashed blue) light polarization and of cubic Si (dashed orange). Our calculation for cubic Si is in good agreement with experimental measurements\cite{PhysRevB.36.4821} (orange dots) apart from a shift in the energy. 4H Si exhibits an optical onset energy significantly lower than cubic Si due to the small direct gap at $\Gamma$.
    (b) Calculated absorption coefficient averaged over light polarizations in the indirect region for 4H (black) and cubic (orange) Si. Experimental measurements for cubic Si from Ref.\cite{PhysRev.109.695} are shown for comparison. A scissor shift of $\Delta = -0.17$ eV is applied to both polytypes to match the band gap of the cubic polytype to experimental measurement of 1.12 eV at room temperature\cite{sze2006o}. The 4H polytype demonstrates stronger indirect absorption than the cubic polytype due to both its smaller indirect gap and the symmetry of its hexagonal structure.}
    \label{fig:optics}
    \vspace{-0.5 cm}
\end{figure}

We next calculate the direct optical absorption spectra for both 4H and cubic Si, both at the independent quasiparticle level and by considering electron-hole interactions with the Bethe-Salpeter equation. We find that excitonic effects are comparable for the two different light polarizations: for the ordinary polarization, a redshift of $\sim$0.12 eV is observed while a similar redshift of $\sim$0.10 eV is observed for the extraordinary polarization due to excitonic effects from both exciton binding and change in oscillation strength. Figure~\ref{fig:optics}(a) shows the calculated spectra for light with ordinary polarization ($\perp c$) and that with the extraordinary polarization ($\parallel c$). We observe that the onset of absorption for the extraordinary polarization is about 0.2 eV higher in energy and sharper than that of the ordinary polarization. This suggests a double onset upon absorption of unpolarized light, which is expected for hexagonal materials especially with relatively long stacking sequences along the $c$ direction. Comparing the cubic polytype and the hexagonal polytype, it can be seen from Fig.~\ref{fig:optics}(a) that overall 4H Si exhibits an absorption onset at a lower photon energy due to its narrower direct gap. Although the absorption onset of 4H Si is less steep compared to that of the cubic Si, the difference in the direct band gap indicates that 4H Si is a much better absorber especially in the visible part of the spectrum of 1.8 eV to 3.1 eV. Such a property makes 4H Si potentially a much better candidate for thin-film solar cells.

%\begin{figure}
%    \centering
%    \includegraphics[width=0.48\textwidth]{figures/fig5.pdf}
%    \caption{Calculated absorption coefficient averaged over light polarizations in the indirect region for 4H (black) and cubic (orange) Si. Experimental measurements for cubic Si from Ref.\cite{PhysRev.109.695} are shown for comparison. A scissor shift of $\Delta = -0.17$ eV is applied to both polytypes to match the band gap of the cubic polytype to experimental measurement of 1.12 eV at room temperature\cite{sze2006o}. The 4H polytype demonstrates stronger indirect absorption than the cubic polytype due to both its smaller indirect gap and the symmetry of its hexagonal structure.}
%    \label{fig:indabs}
%    \vspace{-0.5 cm}
%\end{figure}

While the direct absorption spectra already show the potential of 4H Si in thin-film solar cells, it is also important to accurately examine indirect (phonon-assisted) optical absorption in order to fully quantify the absorption spectrum of solar radiation. This is because the difference of the fundamental indirect gap between the two polytypes is much smaller compared to the difference in direct gaps, while the indirect region is much more important for the application of these materials under the solar spectrum. Figure~\ref{fig:optics}(b) shows the calculated absorption coefficient in the indirect regime for the two polytypes. A rigid shift of -0.17 eV is applied to the onset of both polytypes so that the band gap of the cubic polytype matches the experimental value of 1.12 eV\cite{sze2006o}. In the near IR region, 4H Si is a stronger absorber for both light polarizations, especially in the 1$\sim$1.5 eV spectral region. This is due to two main reasons. First, 4H Si exhibits a narrower indirect gap compared to cubic Si. Although the difference is only 0.05 eV between the two polytypes, this causes a difference of almost one order of magnitude for the absorption coefficient in the indirect region of the spectrum.  Moreover, it can be seen from the Fig. \ref{fig:optics}(b) that the difference between the onset is larger than 0.05 eV, i.e., larger than the band-gap difference, and this is attributed to the stronger electron-phonon interactions in the hexagonal structure.

\begin{figure}
    \centering
    \includegraphics[width=0.5\textwidth]{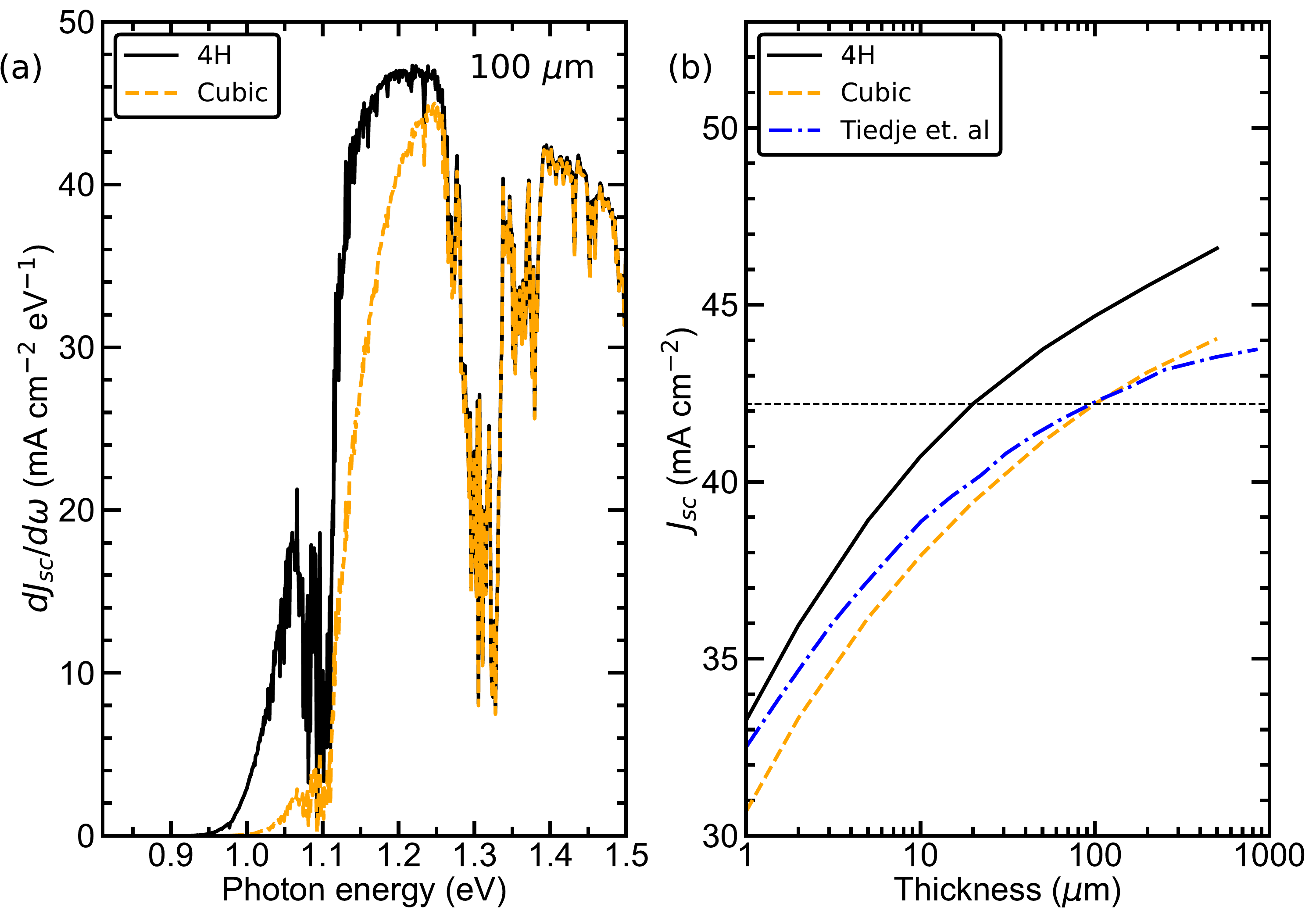}
    \caption{(a) Calculated current density as a function of the photon energy for the AM1.5 solar spectrum\cite{solar} for 100 µm thick textured films of cubic (orange) and 4H Si (black). (b) Current density integrated over photon energies as a function of film thickness for cubic (orange) and 4H (black) Si as well as comparison with the values obtained with the same model from Ref.\cite{tiedje1984} using experimental absorption coefficients (blue). A five times thinner film of 4H Si results in the same current density of 42 mA$\cdot$cm$^{-2}$ as a 100 $\mu$m film of cubic Si, as indicated by the black dashed line. 
    }
    \label{fig:isc}
    \vspace{-0.6 cm}
\end{figure}

To quantify the advantages of the 4H polytype compared to cubic Si in photovoltaic applications, we calculate the photo-induced current density of textured films of various thickness for the two polytypes. We follow the model of Tiedje et. al\cite{tiedje1984} to calculate the current density from photogenerated electron-hole pairs using the AM1.5 solar spectrum and the absorption coefficient of the bulk material. The absorption coefficient of the 4H polytype is averaged over the in-plane and out-of-plane light polarizations to model the absorption for a textured surface. Under ideal conditions, i.e., without considering loss induced by free-carrier absorption, the absorbance ($A$) of a textured film of thickness ($L$) as a function of photon energy ($\omega$) in the weak absorption limit ($\alpha L\ll 1$) is given by:
\begin{equation}
    A(\omega,L)=\frac{\alpha(\omega)}{\alpha(\omega)+1/(4n_r(\omega)^2L)},
\end{equation}
where $\alpha(\omega)$ is the absorption coefficient and $n_{r}(\omega)$ is the photon-energy-dependent refractive index of the material. The term of $(4n_{r}^2(\omega)L)^{-1}$ results from the increased mean path length for light rays through internal scattering due to the assumption of nonspecular textured surface\cite{tiedje1984}. We evaluate $n_{r}(\omega)$ only from the calculated real part of the dielectric function $\text{Re }\varepsilon(\omega) $ via the relationship: $[n_r^2(\omega)=\text{Re }\varepsilon(\omega)]$ because the imaginary part of the dielectric function is small compared to the real part in the range of photon energies where $A(\omega,L)\ll 1$. Following the absorbance, the short-circuit current density per photon energy $dJ_{sc}/d\omega$ is calculated via:
\begin{equation}
    dJ_{sc}/d\omega=eA(\omega, L)n_{ph}(\omega),
\end{equation}
where $n_{ph}(\omega)$ is the photon flux obtained from the AM1.5 spectrum $S(\omega)$ via the relationship: $[n_{ph}(\omega)=S(\omega)/\hbar\omega]$. The integral of $dJ_{sc}/d\omega$ over energy gives the short-circuit current $J_{sc}(L)$ as a function of film thickness. Figure\ref{fig:isc}(a) shows the calculated current density as a function of photon energy for 100 $\mu$m textured thin films of both the 4H and the cubic polytype. We show that for a 4H Si thin film with a thickness of 100 $\mu$m, the short-circuit current density for photon energies in the range of 0.9 eV to 1.2 eV is significantly higher than that of the cubic Si, resulting in 6\% higher integrated short-circuit current density. We also evaluate the current density as a function of film thickness (Fig. \ref{fig:isc}b) in order to quantify the effects of the polytype on solar-cell performance. For the cubic polytype, our calculated $J_{sc}$ agrees well with that calculated with experimentally measured absorption coefficients More importantly, Fig.\ref{fig:isc}b shows that the integrated $J_{sc}$ for the solar spectrum as a function of thickness for 4H Si is about 5\% higher than that of cubic Si. Comparing the two curves horizontally, our calculation indicates that a 20 $\mu$m film of 4H Si achieves a similar short-circuit current density as a five times thicker (100 $\mu$m) film of cubic Si. As a result, the 4H polytype of Si can be a potential excellent candidate for manufacturing thinner silicon-based solar cells.

While our study shows promising potential for 4H Si in thin-film solar-cell applications, the large-scale growth of high-quality 4H Si films has not yet been achieved experimentally. The authors of Ref. \onlinecite{PhysRevLett.126.215701} utilized metastable allotropic transition to synthesize bulk 4H Si by annealing Si$_{24}$ crystals. The authors proposed that the metastable allotropic transition can serve as a way of growing seed 4H Si crystals, which can result in the synthesis of nanostructures and for device-level applications but it remains a challenge to grow high-quality 4H Si using deposition and epitaxial techniques. Therefore, the advantages of 4H Si for optoelectronic applications predicted by our theoretical work aim to stimulate and guide experimental synthesis and characterization of this material for next-generation Si optoelectronic devices. 

In summary, we studied the structural, electronic and optical properties of the 4H Si polytype, and we show that it is a promising solar-cell material. We find that the 4H polytype of Si exhibit a slightly narrower band gap compared to its cubic variation. Due to this difference in the band gap, as well as the stacking along the out-of-plane direction, the optical properties are also affected. We show that the absorption coefficient of 4H Si is much higher than cubic Si across the IR and visible light range. Subsequent calculations on textured thin films demonstrate that application of 4H Si in solar cells can allow up to five times thinner cells compared to the cubic polytype with similar performance, demonstrating its potential in optoelectronic applications.

See supplemental information section I for detailed computational parameters, section II for the comparison of the excitonic effects for 4H silicon for different light polarizations, and section III for the analysis of the contributions of electron-phonon coupling and band gap on the change in the absorption coefficient.

\acknowledgments
%\xz{This needs to be updated. DOE funding added. }
The work is supported as part of the Computational Materials Sciences Program funded by the U.S. Department of Energy, Office of Science, Basic Energy Sciences, under Award No. DE-SC0020129.
Computational resources were provided by the National Energy Research Scientific Computing Center, which is supported by the Office of Science of the U.S. Department of Energy under Contract No. DE-AC02-05CH11231.
%\clearpage
\bibliography{apssamp}

\renewcommand\theequation{S\arabic{equation}}
\renewcommand\thefigure{S\arabic{figure}}

\setcounter{equation}{0} 
\setcounter{figure}{0} 

%\maketitle

\clearpage
\onecolumngrid
\section*{Supplemental information}

\subsection{Computational parameters}
\label{sec:app_comp}

%To obtain relaxed structures for the two polytypes, we apply density functional theory with the Perdew-Burke-Ernzerhof (PBE) exchange-correlation funtional\cite{PhysRevLett.77.3865} using SG15 Optimized Norm-Conserving Vanderbilt (ONCV) pseudopotentials\cite{schlipf2015optimization,PhysRevB.88.085117}.
For the first-principles ground state calculations, the wavefunctions for both polytypes are expanded in terms of plane waves with an energy cutoff of 80 Ry during structural relaxation, and up to 60 Ry for calculations of electronic and optical properties. For both polytypes, optimized structures are obtained by fitting the energy-volume curve to the Murnaghan equation of state\cite{EoS}. The total stress on the cell is less than 10$^{-6}$ Ry/Bohr$^3$ after relaxation. Full relaxation of the unit cell is performed for 4H Si until the Hellmann–Feynman force is smaller than 10$^{-5}$ Ry/Bohr. Relaxation of the structures is performed with Brillouin zone (BZ) sampling grids of $8\times8\times8$ for cubic Si, and $12\times12\times4$ for 4H Si. For both polytypes, the grid is shifted by half a grid spacing to improve convergence. 
To obtain quasiparticle energies, we adopted the single-shot GW approximation as implemented in the BerkeleyGW code\cite{DESLIPPE20121269,hybertsen1986} using BZ sampling grids of $6\times6\times6$ for the cubic and $6\times6\times3$ for the 4H polytype. The static dielectric function was calculated with a 30 Ry plane-wave cutoff and extended to finite frequency using the generalized plasmon-pole model of Hybertsen and Louie\cite{hybertsen1986}. 
To calculate direct optical properties, we solve the Bethe-Salpeter equation for optical polarization functions to take electron-hole Coulomb interactions into account\cite{rohlfing2000o}. 
The electron-hole interaction kernels are calculated using the same BZ sampling grids as in the GW calculations and interpolated to finer grids using the wavefunction overlap between the fine and coarse grids\cite{DESLIPPE20121269}. 
The fine grids used are $12\times12\times12$ for the cubic polytype and $12\times12\times6$ for the 4H polytype with an arbitrary small shift to ensure smooth spectra. 
For indirect optical property calculation, a fine Brillouin zone sampling of $32\times32\times32$ is used for both the electrons and the phonons to ensure smooth spectra.

\clearpage

\subsection{Excitonic effects in 4H silicon}
\label{sec:app_excitonic}
%\begin{figure}
%    \centering
%    \includegraphics[width=0.48\textwidth]{figures/optics_4H_3C.pdf}
%    \caption{Comparison between the calculated direct spectra of 4H Si (blue and black) and cubic Si (red). 
%    The onset of absorption for 4H polytype exhibits clear optical anisotropy, and is 0.6 eV lower than the cubic polytype. 
%    }
%    \label{fig:4h3c}
%\end{figure}
In this section, we show the direct spectra calculated within independent particle approximation and that of including electron-hole interactions via solving the Bethe-Salpeter equation. Including electron-hole interactions results in a redshift of the absorption onset of around 0.1 eV for both the ordinary polarization and extraordinary polarization of light. 

\begin{figure}[!ht]
    \centering
    \includegraphics[width=0.48\textwidth]{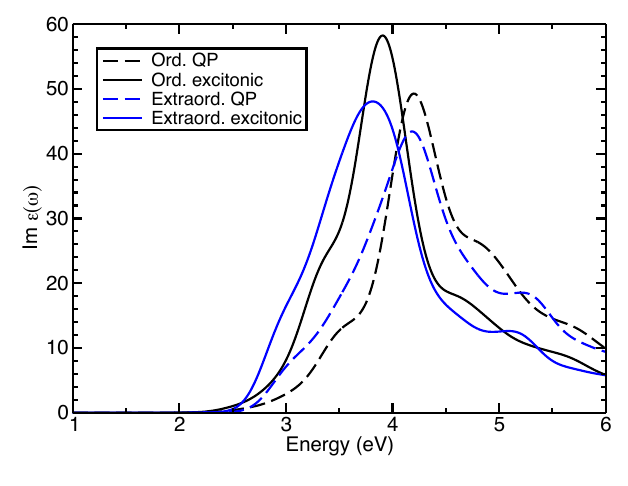}
    \caption{The optical absorption spectra (imaginary dielectric function) of the 4H polytype of Si for different polarizations of light evaluated taking excitonic effects into account via the BSE approach, compared to only consider quasiparticle effects via the GW approximation. Similar excitonic shifts are observed for both the in-plane polarization and the out-of-plane polarization.}
    \label{fig:exciton}
\end{figure}

\clearpage

\subsection{Phonon-assisted spectra assuming the same band gaps of the polytypes}
\label{sec:app_indabs}
\begin{figure}[!ht]
    \centering
    \includegraphics[width=0.48\textwidth]{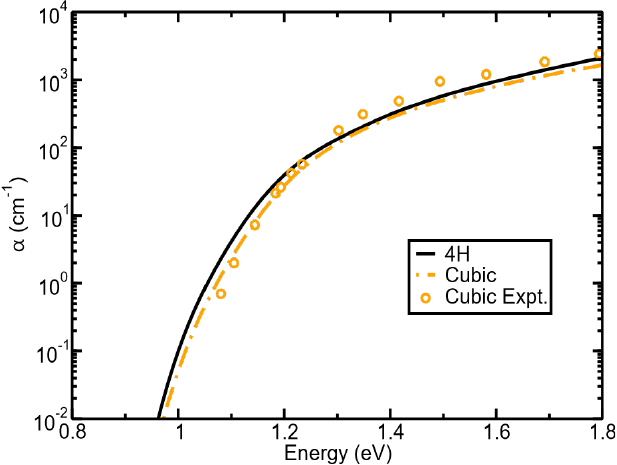}
    \caption{The calculated absorption coefficient in the indirect absorption region for both 4H and cubic Si, with an additional 0.05 eV redshift of the cubic spectrum to account for the band gap difference between the two polytypes. Under this assumption, the 4H polytype still shows a larger absorption coefficient across the energy range considered.
    }
    \label{fig:comp_shift}
\end{figure}

In this section, we illustrate that the stronger absorption of 4H Si is due to not only the band gap, but also stronger electron-phonon interactions. In Fig.\ref{fig:comp_shift}, we show the absorption coefficient of 4H Si and cubic Si without considering the difference in the band gap. 
In the figure, instead of a $\Delta=-0.17$ eV scissor shift for both polytype as mentioned in the main text, we apply a scissor shift of $\Delta=0.17$ eV to the cubic polytype, and $\Delta=-0.22$ eV to the 4H polytype so the difference in the band gap is compensated. 
As can be seen from the figure, the absorption coefficient of 4H Si is still higher than that of the cubic polytype by a maximum factor of two around the 1 eV to 1.3 eV region. This difference is caused by the stronger electron-phonon interaction due to the hexagonal structure itself.

% Produces the bibliography via BibTeX.

\end{document}